\documentclass[prb,twocolumn,showpacs,amsmath,amssymb,floatfix]{revtex4}

\usepackage{times}

\begin{document}

\def\SAVE#1 {{}}   
\def\MEMO#1 {{}}   
\def\secc#1{\vskip 0.1 truein {\underline{\it #1}} ---}

\def \be {\begin{equation}}
\def \eeq {\end{equation}}

\newcommand{\dom} {{\rm dom}}
\newcommand{\var} {{\rm var}}
\newcommand{\ehat} {{\bf \hat e}}
\newcommand{\eperp} {{\ehat^\perp}}
\def \hcolor {{\bf h}}
\def \Fcolor {{F_\hcolor}}
\def \Kcolor {{K}}
\def \Kcrit {{K_c}}
\def \Fvar {{F_\var}}
\newcommand{\HH} {{\cal H}}
\def \tHH {{\tilde{\HH}}}
\newcommand{\HHtwo} {{\HH^{(2)}}}
\newcommand{\HHthree} {{\HH^{(3)}}}
\newcommand{\HHfour} {{\HH^{(4)}}}
\newcommand{\tHHquart} {{\tHH^{(4)}}}
\newcommand{\HHQ} {{\cal Q}}
\newcommand{\HHPotts}{{\Phi}}
\newcommand{\HHsw} {{\HH_{\rm sw}}}
\newcommand{\tHHQ} {{\tilde{\cal F}_{\rm soft}}}
\newcommand{\ZC} {{\cal Z}}
\def \Jcal {{\cal J}}
\def \q {{\bf q}}
\def \qq {{\bf q}}
\def \rr {{\bf r}}
\def \RRhat {{\bf \hat R}}
\def \uu {{\bf u}}
\def \nn {\hat{\bf n}}
\def \zz {\hat{\bf z}}
\def \ss {{\bf s}}
\def \la {{\langle}}
\def \ra {{\rangle}}
\def \half{{\frac{1}{2}}}
\def \ising {{\eta}}
\def \col{{c}}
\def \eqr#1{(\ref{#1})}

\title{Long range order in the classical kagome antiferromagnet:
effective Hamiltonian approach}

\author{Christopher L. Henley}
\affiliation{Laboratory of Atomic and Solid State Physics, Cornell University,
Ithaca, New York, 14853-2501}

%%% \date{\today}

\begin{abstract}
\MEMO{May be over 600 char., but is anyone checking exactly?}
Following Huse and Rutenberg [Phys. Rev. B 45, 7536 (1992)], 
I argue the classical Heisenberg antiferromagnet
on the kagom\'e lattice has long-range spin order
of the $\sqrt{3}\times\sqrt{3}$ type in the 
limit of zero temperature.
%% (modulo gradual orientation fluctuations of the spins' plane).
I start from the effective quartic Hamiltonian for the
soft (out of plane) spin fluctuation modes, and treat as 
a perturbation those terms which depend on the discrete
coplanar state.  Soft mode  expectations
become the coefficients of a discrete effective Hamiltonian, 
which (after a coarse graining) has the sign favoring
a locking transition in the interface representation of the
discrete model.
%%% using a ``Coulomb phase'' coarse-graining.
\end{abstract}

%%% {75.10.Hk=Classical spin models,
%%%  75.50.Ee=Antiferromagnetics,
%%%  75.40.Cx=Static properties}
\maketitle

\MEMO{ New version, 10/9/09, is c 20 lines overlong.
Cuts can be achieved by shortening the section
on asympotic behavior.}

\section{introduction}
\label{sec:intro}

Consider the nearest-neighbor antiferromagnet with 
classical spins of $n=3$ components on the kagom\'e 
lattice of corner-sharing triangles,
  \be
      \HH = J \sum _{\la ij \ra} \ss_i\cdot \ss_j .
  \eeq  
This is the prototypical {\it highly frustrated} system,  
meaning its ground state manifold has macroscopically
many degrees of freedom, and any spin order or freezing
sets in at temperatures $T \ll J$~\cite{ramirez-hfm}.
It is well established that as $T\to 0$,
the free energy of spin-mode fluctuations 
causes ordering into a {\it coplanar} state, 
a particular kind of classical ground state in which all spins 
lie in the same plane of spin space pointing
in just three directions ($120^\circ$ apart)~\cite{chalker}.
These directions -- which can be written as colors
$c_i\equiv A$, $B$, or $C$ taken by spins in a 3-state Potts model -- 
constitute a 3-coloring (the ground state constraint
implies every triangle has one of each color).
The number of such colorings is exponential in the system size.
The same is true for three-dimensional lattices
of corner-sharing triangles such as the (half) garnet lattice~\cite{petrenko}
or equivalently hyperkagom\'e lattice~\cite{hyperkagome},
and others~\cite{henley-HFM08}.

\SAVE{I said the ordering into a coplanar state is at
harmonic order; Canals points out Moessner and Chalker
attribute it to counting quartic modes.  My focus is
on whether one breaks the degeneracy; whether this can
cause order is a separate picture.}

Can the coloring achieve a long-range order?
All simulations~\cite{chalker,reimers,zhitomirsky,robert} 
indicate the Potts spins are disordered 
(or algebraically correlated) as
in the unweighted coloring (see below).
However, following Huse and Rutenberg~\cite{huse}, I propose
this coloring develops long-range order in
the $T\to 0$ limit,
as a consequence of the {\it unequal} weighting of the 
discrete states, when one takes into account the free
energy of fluctuations about each state.
Of course, for $d=2$ at $T>0$, the orientation of the spin plane 
must fluctuate slowly in space; 
%%%%%%%%%%%%%%%%%%%%%%%%%%
\SAVE{I think the correlation length scales as 
$\xi_{\rm plane} \sim \exp({\rm const}/\sqrt{T})$, 
from plugging in a bending stiffness $\rho= \sim \sqrt{T}$ 
into the usual formulas $\exp (\rho/T)$ for e.g. the
Heisenberg model.  However, I didn't find this anywhere
in the literature.}
%%%%%%%%%%%%%%%%%%%%%%%%%%
nevertheless the colors/Potts directions may be unambiguously defined 
throughout the system.  
%% We might consider them to be a sort of ``hidden order parameter''; alternatively
But my goal is only the $T\to 0$ limiting ensemble, 
well defined (and nontrivial) since the obtained effective Hamiltonian 
[e.g. \eqr{eq:J-G-Gamma}] scales as $T$;
whereas the spin-plane correlation length diverges exponentially~\cite{FN-vortices}
as $T\to 0$.

The calculation entails a series of mappings and effective Hamiltonians.
First I shall review how, starting from the usual spin-deviation
expansion, one integrates out most of the fluctuations
leaving a {\it quartic} effective Hamiltonian $\HHQ$ for the 
dominant fluctuations~\cite{shender}.
The largest terms of $\HHQ$ are independent of the discrete
Potts configuration, so treating the rest as a perturbation
yields an effective Hamiltonian $\HHPotts$ for the Potts spins, 
purely entropic in that $\HHPotts \propto T$.
Its coefficients may be inferred from simulations, or
approximated analytically (taking advantage of a ``divergence constraint'' 
on the dominant fluctuations).
The Potts spins map in turn to a ``height model'',
whence it becomes clear that $\HHPotts$ causes locking into an
ordered state~\cite{huse}.
The expected long-range order 
is too tenuous to see directly in simulations, but
might be estimated analytically from the height model.

\section{Effective Hamiltonian derivation}
%%% ... and perturbation expansion
\label{sec:Heff}
%%% \secc{Deriving effective Hamiltonian}
%%%%%%%%%%%%%%%%%%%
The object is to obtain an effective Hamiltonian for any
of the discrete coplanar ground states, which absorbs the 
free energy of the low-temperature (anharmonic) fluctuations 
about that state.
%%%%
The first step is the ``spin-wave'' expansion in deviations
from a given coplanar state.  
We parametrize the out-of-plane deviation as $\sigma_i$,
and the other deviation component as $\theta_i$, 
the spin's in-plane rotation about the plane normal axis.
%%%%%%%%%%%%%%%%%
\SAVE{Ref.~\onlinecite{chalker},
defined an orthonormal triad at each site
$(\hat {\bf x}_i,\hat {\bf y}_i, \hat {\bf z}_i)$,
such that $\hat {\bf x}_i$ is the spin's classical direction 
and $\hat{\bf y}_i$ is the direction normal to the plane of coplanarity.
In that notation, the $y$ spin deviation is $\sigma_i$ while the $z$ deviation
is parametrized by $\theta_i$, the spin's in-plane rotation 
away from the $\hat{\bf x}$  axis.}
%%%%%%%%%%%%%%%%%%%%%%%%%%%%%%%%%%
Then the spin-wave expansion (I set $J\equiv 1$) is
  \begin{subequations}
  \label{eq:HHsw}
  \begin{eqnarray}
     \HHsw &\equiv& \HHtwo + \HHthree + \HHfour + \ldots,\nonumber \\
     \HHtwo &=& 
     \sum _{\la ij\ra}
          \Big[ \frac{1}{4}(\theta_i-\theta_j)^2  +  \sigma _i \sigma _j \Big] +
        \sum _i \sigma_i^2 ;
      \label{eq:h2} \\
     \HHthree_\dom &=&  \sum _\alpha  \eta_\alpha \HH^{(3,\alpha)}_\dom, 
      \label{eq:h3sum} \\
    \HHfour_\dom &=&  \sum _{\la ij \ra}
       \frac{1}{16} (\sigma_i^2  - \sigma_j^2 )^2 .
   \label{eq:h4}
  \end{eqnarray}
  \end{subequations}
In Eq.~\eqr{eq:h3sum}, $\alpha$ indexes the center of each triangle, and
   \be
      \HH^{(3,\alpha)}_\dom \equiv  
      \frac{\sqrt 3}{4}
%%%      \frac{1}{2} \sin \frac{2 \pi}{3} \;
       \sum _{m=1}^3 \big[\sigma_{\alpha m}^2
            (\theta_{\alpha,m+1}-\theta_{\alpha, m-1})\big].
      \label{eq:h3}  
    \eeq
From here on, I use ``$\alpha m$'' ($m=1,2,3$)
to denote the site on triangle $\alpha$ in sublattice $m$,
as an alias for the site index ``$i$''; the index $m$
is taken modulo 3  (in expressions like ``$m+1$'') 
and runs counterclockwise around the triangles 
whose centers are even sites on the honeycomb lattice
of triangle centers.
Following Ref.~\onlinecite{shender},
I retained only ``dominant'' anharmonic terms 
$\HHthree_\dom$ and $\HHfour_\dom$,
being the parts of \eqr{eq:h3} and \eqr{eq:h4}
containing the highest powers of $\sigma$
(this will be justified shortly).

The $\eta_\alpha$ prefactor in $\HHthree$
is the {\it only} dependence in Eq.~\ref{eq:HHsw} on the coloring state;
this ``chirality'' $\eta_\alpha$,
is defined by $\eta_\alpha \equiv +1(-1)$  when the Potts labels 
are ordered as  ABC (CBA) as one walks counterclockwise 
about triangle $\alpha$.
It is convenient to  label coplanar states by the configuration $\{ \eta_\alpha \}$
~\cite{FN-not-1to1}.
%%%%%
Then a discrete Hamiltonian $\HHPotts$  can be defined for colorings,
a function of the $\{ c_i \}$ implicitly through the $\eta_\alpha$'s
in \eqr{eq:HHsw}:
   \be
       e^{-\HHPotts(\{\col_i\}/T} = 
       {\ZC}(\{ \col_i\}) \equiv
       \int_{\rm basin} \prod_i (d\theta_i \, d\sigma_i) e^{-\HHsw/T}
    \label{eq:HHPotts}
    \eeq
As $T\to 0$, the ensemble weight concentrates closer
and closer to the coplanar state~\cite{chalker,moessner};
the integral in \eqr{eq:HHPotts} is implicitly limited to
the ``basin'' in configuration space centered on one coplanar
state, and $\ZC(\{ \col_i \})$ is the portion of the total
partition function assigned to 
the corresponding coloring.
Since $\HHtwo$ is independent of which coplanar state we are in,
$\HHPotts$ is independent of $\{ \col_i \}$ at harmonic order.

\SAVE{The barrier height defining a basin scales as T ln T, 
I believe, so the suppression of the barrier states goes to zero 
only as a power law in $T$, not exponentially.  Since the coplanarity
potential is purely entropic, the barrier value scales entirely
as T.  But the lower cutoff goes lower, that is what (I believe)
gives the logarithm.}
 
Before we go on to anharmonic order, let's note 
the $\sigma$ part of $\HHtwo$ can be written
$\HHtwo_\sigma = \half \sum _\alpha (\sum _m \sigma_{\alpha m})^2$.
So there is a well-known whole
branch of out-of-plane ($\sigma$) modes, called ``soft modes'',
having {\it zero} cost at harmonic order; the soft mode subspace is
defined by the constraint
   \be
      \sum _{m=1}^3 \sigma_{\alpha m} = 0   \qquad ({\rm soft})
   \label{eq:soft-sum}
   \eeq
being satisfied on every triangle $\alpha$.  
(Two more out-of-plane branches, 
as well as all $\theta$ branches, are called ``ordinary'' modes.)
Being limited only by higher order terms, 
soft modes have large mean-square fluctuations, of $O(\sqrt T)$, 
compared to $O(T)$ for ordinary modes~\cite{chalker,shender}; 
this explains why factors containing soft modes were
``dominant'' in Eq.~\eqr{eq:HHsw}.
The $\sigma_i$'s in ``dominant'' terms are
limited to the ``soft'' subspace satisfying \eqr{eq:soft-sum}.
%%%%%%%%%%%%%%%%%%%%
\SAVE{Interestingly, all dependences on soft fluctuations are through 
square factors $\sigma_i^2$.}

The next step is to do the Gaussian integral over all 
$\theta_i$ modes~\cite{chalker},
as worked out in Ref.~\onlinecite{shender},
obtaining a quartic effective Hamiltonian $\HHQ$ 
for only soft modes:
\be
     \HHQ = \HHfour_\dom
     - \sum_{\alpha,\beta} 
   \eta_\alpha \eta_\beta \HHQ'_{\alpha\beta}
     \label{eq:quartic-soft}  
\eeq
with~\cite{FN-sigma4-deception}
   \be
     \HHQ'_{\alpha,\beta}  \equiv \sum _{m,n=1}^3
   \big(\frac{\sqrt 3}{4}\big)^2
%%%     \frac{1}{4} \sin^2 \frac{2\pi}{3}\;
     G_{\alpha m, \beta n} \sigma_{\alpha m}^2 \sigma_{\beta n}^2
    \label{eq:quartic-pair}
   \eeq
The Green's function of the $\theta$ modes 
%%% being integrated out 
was defined by
    \be
           T\; G_{\alpha m, \beta n} \equiv
       \la (\theta _{\alpha, m+1}-\theta_{\alpha, m-1})
        (\theta _{\beta, n+1}-\theta_{\beta, n-1})\ra _\theta
    \label{eq:Gdef}
     \eeq
where ``$\la ... \ra_\theta$'' means taken in the (Gaussian)
ensemble of $\HHtwo_\theta$ ($\equiv$ the  $\theta$ part of $\HHtwo$).
As $G_{ij}$ decays with distance, the largest terms are
state-independent:
$\HHQ'_{\alpha\alpha} = (3/16)[G_0\sum _{m=1}^3 \sigma_{\alpha m}^4 +
  2 G_1 \sum _{m<n} \sigma_{\alpha m}^2 \sigma_{\alpha n}^2].$, 
where $G_0$ and $G_1$ are the on-site and first-neighbor $G_{ij}$.
Trivially $2G_1=-G_0$, and $G_0\equiv 1$ (due to equipartition,
which implies $\la \HHtwo_\theta \ra = 3T/4$ per triangle).
Also, given \eqr{eq:soft-sum}, 
%%% there is only one symmetric quartic polynomial within each triangle, 
$\sum _{m<n} \sigma_{\alpha _m}^2 \sigma_{\alpha n}^2 \to 
\half \sum _m \sigma_{\alpha m}^4$
%%%%%%%%%%%%%%%%%%%%%%%%%%%%%%
\SAVE{I got an estimate for an isolated triangle 
(as in a Husimi cactus), $G_0 =2/3$ exactly.
But much better, we can use equipartition to show
$G_0=2/z$, where $z$ is coordination, i.e. $G_0=1/2$ exactly.}
%%%%%%%%%%%%%%%%%%%%%%%%%%%%%%
in $\HHQ'_{\alpha\alpha}$, and similarly in \eqr{eq:h4}
$\HHfour_\dom \to (1/16) \sum _i \sigma_i^4$.
Finally we can regroup \eqr{eq:quartic-soft}  as
  \be
          \HHQ = \HHQ_0  -\sum _{\alpha\neq \beta}
      \eta_\alpha \eta_\beta \HHQ'_{\alpha\beta},
      \HHQ_0 =  B_0 \sum _i \sigma_i^4
   \label{eq:HHQ-local}
   \eeq
with $B_0 =13/16$
from both $\HHfour_\dom$ and $\HHQ'_{\alpha\alpha}$ terms.
%%%% B_0 \equiv \frac{1}{4} (1 + \frac{9}{4} G_0)=\frac{13}{16}$.

%%% \secc{Perturbation expansion}
%%%%%%%%%%%%%%%%%%%%%
Now I turn to the perturbation expansion:
the key step in our whole derivation is to 
expand \eqr{eq:HHPotts}
treating the $\{ \eta  _\alpha \}$
as if they were small quantities.
(In fact $|\eta_\alpha| =1$, so a 
perturbative treatment might appear questionable,
but quantitatively $\HHQ_0$ has 
a much larger coefficient than the terms in $\HHQ'$,
owing to the decay of $G_{ij}$ with separation.)
\MEMO{TODO: I MUST CHECK the values of those coefficients
(for second or third neighbors).}
%%%%%%%%%%%%%%%%%%%%%%%%%%%%%%%%%%%%%%%
The resulting (and final) effective Hamiltonian is, 
to lowest order,
   \be
      \HHPotts  = - \frac{1}{2} \sum_{\alpha\neq\beta}
     \Jcal_{\alpha\beta} \eta _\alpha \eta _\beta
   \label{eq:HPotts-pair}
   \eeq
with 
    \be
     \Jcal_{\alpha\beta} \equiv \la \HHQ'_{\alpha\beta} \ra_0.
     = \sum _{m,n=1}^3
   \big(\frac{\sqrt 3}{4}\big)^2
     G_{\alpha m, \beta n} 
       \la \sigma_{\alpha m}^2 \sigma_{\beta n}^2 \ra_0
    \label{eq:Jresult}
    \eeq
where the expectation is taken in the ensemble of $\HHQ_0$.
Notice that since $\HHQ$ is homogeneous in $\{\sigma _i \}$,
it follows that
the partial partition function $\ZC(\{ \col_i\})$
in \eqr{eq:HHPotts} -- and consequently $\HHPotts/T$ --
is {\it temperature independent} as $T\to 0$, 
apart from a configuration-independent powers of $T$.

\SAVE{ $T^{1/4}$ per mode.
That $T^{1/4}$ factor was responsible for the anomalous 
specific heat of $k_B/4$ per soft mode~\cite{chalker}.}

\SAVE{A similarly brutal approximation 
turned out to be justified in an analogous calculation for 
an effective Hamiltonian of the large-$S$ quantum antiferromagnet
on the pyrochlore lattice~\cite{pyrochlore}.}

A corollary of my assumption that $\HHQ'_{\alpha\beta}$ 
is ``small'' is that expectations $\la ... \ra_{\rm sw}$
of polynomials in $\{ \sigma_i \}$,
measured under the {\it full} spin-wave Hamiltonian
$\HHsw$, 
should be practically independent of the coloring configuration 
$\{ \eta_\alpha \}$~\cite{FN-Goldstone}.
That can be checked in Monte Carlo or molecular dynamics 
simulations~\cite{FN-higher-order}
of the Heisenberg model.  The needed correlations can be measured
even if the system is confined to the ``basin'' of one coplanar state: 
there is no need to equilibrate the relative occupation 
of different basins.
\SAVE{Which may require activation over barriers;
however, Robert and Canals suggest that -- when MD is used --
the system rather rapidly transits between basins.}
%%%%%%%%%%%%%%%%%%%%%%%%
Those same simulations would 
numerically evaluate the quartic expectations
%%%%%%%%%%%%%%%%%%%%%%%%%%%%%%
\SAVE {Quartic expectation could also
be checked against the variational predictions (below)
for the angle and distance dependence, and 
from them the couplings $\Jcal_{\alpha\beta}$. }

\section{Self-consistent approximation for couplings  
   and asymptotic behavior}
\label{sec:self-cons}
%%% \secc{Self-consistent approximation for couplings $\{ \Jcal_{\alpha\beta} \}$}
%%%%%%%%%%%%%%%%%%%%%%%%%%%%%
An alternative to simulation is to analytically
evaluate the quartic expectations in \eqr{eq:Jresult}.
using a self-consistent decoupling.
That is,  \eqr{eq:HHQ-local}  is replaced by 
%%%% $B_0 \sigma _i^4 \to 3 B_0 \la \sigma_i\ra_{\rm var} \sigma_i^2$, 
%%% One way to rationalize this is a a maximum-likelihood viewpoint:  
   \be 
      \Fvar \equiv 
    \half B \sum _i \sigma_i^2, 
   \label{eq:Fvar}
    \eeq
defining a Gaussian variational approximation
to the soft mode ensemble; here
   \be
   B\equiv 6 B_0 \langle \sigma_i^2\ra_\var ,
%%%        B\equiv 6 B_0 \langle |\phi_\mu-\phi_\nu|^2 \ra_\var .
   \label{eq:B}
   \eeq
with ``$\la ...\ra_{\rm var}$'' taken in the ensemble of \eqr{eq:Fvar}.
%%%%%%%%%%%%%%
Now let $\Gamma_{ij}$ (also written $\Gamma_{\alpha m, \beta n}$)
be the Green's function for $\sigma_i$ modes:
  \be
     \la \sigma_i \sigma _j \ra_{\rm var}  = T \; \Gamma_{ij}/B
  \label{eq:Gamma}
  \eeq
(this definition makes $\Gamma_{ij}$ independent of $B$ and $T$)
and let $\Gamma_{ii} \equiv \Gamma_0 = 1/3$.
\SAVE{I got a rough  rough guess of $0.4$, e.g. exactly doing 
a $2\times 2$ cluster of triangles gives $4/9$.  But
equipartition gives us $2/z=1/3$ for a triangular lattice
of hexagon centers.}
%%%%%%%%%%%%%%%%%%%%%%%%%%%%%%%%%%%
Combining \eqr{eq:B} and \eqr{eq:Gamma}, I get
the  self-consistency condition $B = (6 B_0 \Gamma_0 T)^{1/2}=
(13T/8)^{1/2}$.
%%%   \label{eq:self-con}
Next, the expectations in \eqr{eq:Jresult} are evaluated in
the variational approximation, decoupling by Wick's theorem as
   \be
       \la \sigma_i^2 \sigma_j^2 \ra_\var =
       \la \sigma_i^2 \ra_\var \la \sigma_j^2 \ra_\var +
       2 \la \sigma_i \sigma_j \ra_\var^2 =
        \Big(\frac{T}{B}\Big)^2 
         \big[\Gamma_0^2 + 2 \Gamma_{ij}^2 \big] .
   \label{eq:sigma-decoupled}
   \eeq
Substituting \eqr{eq:sigma-decoupled} into \eqr{eq:Jresult} gives
my central result for the effective Hamiltonian,
   \be
      \frac{\Jcal_{\alpha\beta}}{T} \approx \frac{3}{13} 
             \sum _{m,n=1}^3
             G_{\alpha m, \beta n} \Gamma_{\alpha m, \beta n}^2 
   \label{eq:J-G-Gamma}
   \eeq
[$\Gamma_0^2$ from \eqr{eq:sigma-decoupled} always cancels in the $m,n$ sum.] 

Eq.~\eqr{eq:J-G-Gamma} gives
$\Jcal_1/T \approx -1.88 \times 10^{-3}$ and 
$\Jcal_2/T \approx -4.3 \times 10^{-4}$.
Assuming these two dominate, the state with the lowest $\Phi$ value is the ``$\sqrt{3}\times\sqrt{3}$''
pattern, the ``antiferromagnetic'' arrangement of chiralities $\eta_\alpha$, 
but other coloring configurations are only slightly less likely.
The long range order suggested by Ref.~\onlinecite{huse} is a subtle
crossover of correlation functions at large (but not diverging) scales,
best expressed in terms of a ``height model'', as will be
developed in Sec.~\ref{sec:height}.
%%%%  (and physically dependent on the constraint which allows it)

%%% \secc{Asymptotic behavior of couplings}
%%%%%%%%%%%%%%%%%%%%%%%%%%%%%%%
Before that, in order to check that more distant couplings 
$\Jcal_{\alpha\beta}$ can be neglected, 
I will work out how they scaling at large $R$.
We need both kinds of Greens function in \eqr{eq:J-G-Gamma}, 
tackling the  $\theta_i$ fluctuations first.
In Eqs.~\eqr{eq:h2} and \eqr{eq:Gdef},
$(\theta_{m+1}-\theta_{m-1}) 
\approx - a\,\epsilon_\alpha \eperp_m\cdot~\nabla~\theta$,
where $a$ is the nearest neighbor distance, 
and $\epsilon_\alpha = +1 (-1)$ when $\alpha$ labels an 
even (odd) triangle. 
The unit vector $\ehat_m \equiv (\cos \psi_m, \sin \psi_m)$,
is defined to point from the center of any even triangle to 
its $m$  corner, and $\eperp_m \equiv {\bf \hat z} \times \ehat_m$.
At long wavelengths, 
   \be
       \HHtwo_\theta \approx \half \rho_\theta 
          \int d^2\rr |\nabla \theta (\rr)|^2
   \label{eq:Htheta-grad}
   \eeq
where $\rho_\theta=\sqrt{3}/2$. 
%%%%%%%%%%%%%%%%%%
Asymptotically the Greens function of \eqr{eq:Htheta-grad}
is pseudo-dipolar:
   \be
%%%    \begin{eqnarray}
%%%      G_{\alpha m, \beta n} &\approx& \frac{a^2}{2\pi \rho_\theta}
%%%         \epsilon_\alpha \epsilon_\beta \Big( 
%%%       \frac{\eperp_m \cdot \eperp _n -2 
%%%         (\RRhat \cdot \eperp_m) (\RRhat \cdot \eperp_n)}
%%%              {R^2} \Big) \nonumber \\
      G_{\alpha m, \beta n} \approx 
       \frac{a^2}{2\pi \rho_\theta R^2} 
\epsilon_\alpha \epsilon_\beta
    \cos (\psi_m + \psi_n-2 \psi_R) .
    \label{eq:Gdipolar}
    \eeq
Here ($R,\psi_R$)  are the polar coordinates of the 
vector between triangle centers $\alpha$ and $\beta$.

The $\sigma_i$ fluctuations are handled similarly.
The soft-mode constraint \eqr{eq:soft-sum} is
implemented by writing $\sigma_i$ as a discrete gradient,
$\sigma_i \equiv \phi_\nu-\phi_\mu$, analogous
to the ``height'' model constructions~\cite{shender-origami}.
Here $\{ \phi _\mu \}$
is defined on the hexagon centers, and $\mu \to \nu$ is oriented
counter-clockwise around even kagom\'e triangles.  
The discrete gradient defining $\sigma_i$ can
be converted into a continuous one, 
%%% $\rr_\nu-\rr_\mu = 2 a\eperp_m$ and 
$\sigma_{\alpha m} \approx 2 a\, \eperp_m\cdot \nabla \phi$.
Then the long-wavelength limit of \eqr{eq:Fvar} is
looks like \eqr{eq:Htheta-grad}.
with $\rho_\theta \to \rho_\phi =  2 \sqrt 3 B$.
%%%%%%%%%%%%%%%%%%%%%%%%%%%%%%%%
\SAVE{
The equation reads
    \be
    \Fvar = \half B \sum _{\la \mu\nu\ra} |\phi_\mu-\phi_\nu|^2 
          \approx  \half \rho_\phi \int d^2 \rr |\nabla \phi(\rr)|^2 
   %%% \label{eq:F_phi_grad}
   \eeq
     }
%%%%%%%%%%%%%%%%%%%%%%%%%%%%%%%%
That implies that for large separations $R$,
%%%%%%%%%%%%%%%%%%%%%%%%%%%%%%%%
\SAVE{
   \be
     \Gamma_{\alpha m, \beta n} \approx \frac{(2a)^2}{2\pi \rho_\phi R^2} 
     \cos(\psi_m + \psi_n - 2 \psi_R),
   \label{eq:Gamma-far}
   \eeq
  }
%%%%%%%%%%%%%%%%%%%%%%%%%%%%%%%%
$\Gamma_{\alpha m, \beta n}$
looks like Eq.~\eqr{eq:Gdipolar} with $\rho_\theta \to \rho_\phi/4$.
Inserting both Green's function behaviors 
%%% \eqr{eq:Gdipolar} and \eqr{eq:Gamma-far} 
into \eqr{eq:J-G-Gamma}, 
I get the asymptotic behavior of the couplings:
%%% \eqr{eq:Jresult}:
  \be
   \frac{\Jcal_{\alpha\beta}}{T} \approx 
       \frac{A}{(R/a)^6} \epsilon_\alpha\epsilon_\beta \cos 6\psi_R 
  \label{eq:Jasymp}
  \eeq
for large $R$ with $A= {6 \sqrt{3}}/{ 13^2  \pi^3} \approx 2.0\times 10^{-3}$.
%%%%%%%%%%%%%%%%%%%%%%
\SAVE{This came from $\Jcal_{\alpha\beta} \approx 
       \frac{4A}{9R^6} \epsilon_\alpha\epsilon_\beta 
       \sum _{mn} \cos^3 (\psi_m + \psi_n-2 \psi_R)$.
In \eqr{eq:Jasymp}
  \be
       A \equiv  
         \frac{3}{13} \cdot \frac{16T} {(2\pi)^3 \rho_\theta \rho_\phi^2} 
             \cdot \frac{9}{4} = 
         \frac{6 \sqrt{3}} { 13^2  \pi^3} .
   \eeq
}
%%%%%%%%%%%%%%%%%%%%%%
Eq.~\eqr{eq:Jasymp} shows the interaction decays rapidly with distance and 
oscillates as a function of angle.   
%%% In particular, the nearest- and second-neighbor interactions
%%% satisfy $\Jcal_1 <0$ and $\Jcal_2 >0$. 

\SAVE{Besides the large-$N$ way, mentioned in text, there
is one more way to justify 
the quadratic variational approximation:
the ``maximum likelihood'' viewpoint
(this was CLH's viewpoint in Ref.~\onlinecite{pseudodipolar})
The quartic Hamiltonian is purely local apart from the
constraint, so implement it in a minimal fashion.
A second way is RG.  Imagine it was $|\nabla \phi|^2$
and add a quartic correction; that would be irrelevant
in the renormalization group for a Coulomb gas.~\cite{nelson}.}

\SAVE{Whereas the local variable in most discrete height models 
includes a periodic dependence on the coarse-grained height
field, in the present case $\phi_\mu$ is continuous and
there is no such dependence.  So, we only
get the generic correlations due to the divergence constraint,
which have the pseudo-dipolar form.}

\section{Height model and long range order}
\label{sec:height}
%%% \secc{Height model and locking}
%%%%%%%%%%%%%%%%%%%%%%%%%%%%%%%%%
The {\it discrete} ensemble in which all 3-colorings $\{ c_i \}$
are equally likely is known to have power-law correlations,
which may be understood via a mapping of the Potts 
microstates to a two component ``height'' variable
$\hcolor(\rr)$~\cite{huse,kondev}.
At coarse-grained scales, the ensemble
weight of $\{ \hcolor(\rr) \}$ is described by a free
energy 
   \be
       \Fcolor = \int d^2\rr \half \Kcolor |\nabla \hcolor |^2, 
   \label{eq:F-height}
   \eeq
handled by standard Coulomb-gas techniques~\cite{nienhuis}.

Ref.~\onlinecite{huse} pointed out the equal-weighted
coloring has a height stiffness $\Kcolor=\Kcrit$ exactly, where
$\Kcrit$ is the critical value for the roughening transition. 
Any increase in $\Kcolor$ must cause $\hcolor(\rr)$ to lock 
to a uniform mean value.~\cite{nienhuis,nelson,huse}.  
That corresponds to long-range order of the colors ($=$ Potts spins), 
into the pattern of with the flattest $\hcolor(\rr)$, namely the 
``$\sqrt 3 \times \sqrt 3$''  state.  Since (as shown above)
$\HHPotts$ favors that flat state, the coloring ensemble with the
$\HHPotts$ weighting is coarse-grained to a height ensemble 
with a slightly larger $\Kcolor$, and therefore
we get long range order, as claimed.~\cite{huse} 

The couplings $\Jcal_{\alpha\beta}$  as approximated analytically,
or obtained from a simulation, may be used as a Hamiltonian in 
discrete simulations of the coloring model.
These are far faster than simulations of the Heisenberg spins,
but I still doubt such simulations will see long-range order directly, 
in the accessible system sizes.
But the height stiffness $\Kcolor$ can be accurately measured 
(using Fourier transforms~\cite{kondev-sim}.)
With that, by iteration of renormalization-group equations~\cite{nelson},
it should be possible to semi-analytically estimate the 
length scale $\xi$ at which the color correlations cross over
from power-law decay to long-range order, and the size of
the order parameter.

%%% \secc {Extending to $d=3$}
%%%%%%%%%%%%%%%%%%%%%%%%%%%%%%%%%%%%
What happens to this whole story in $d=3$, for the
Heisenberg antiferromagnet on
triangle-sharing lattices~\cite{petrenko,hyperkagome,henley-HFM08}?
A minor difference is that in $d=3$ the spin plane orientation 
has {\it true} long-range coplanar order at some $T>0$, as do
the three spin directions within the plane~\cite{zhitomirsky}.
The derivation and result for the effective  Hamiltonian
\eqr{eq:J-G-Gamma} extend to $d=3$;
%%%%%%%%%%%%%%%%%%%
\SAVE{
The coplanar states are described by the same colorings
and chiralities~\cite{henley-HFM08}; the spin-wave expansion \eqr{eq:HHsw}
and perturbation expansion \eqr{eq:Jresult} look identical
(except the convention for sublattice indices $m$ is less obvious).}
%%%%%%%%%%%%%%%%%%%
There is also the unimportant difference is that, in deriving the asymptotic 
behavior of $\Jcal_{\alpha\beta}$, a 
``Coulomb phase''~\cite{pseudodipolar,isakov}
rather than a ``height function'' viewpoint must be
used for coarse-graining $\sigma$, but
$\Gamma_{ij}$ still  has a pseudodipolar form~\cite{pseudodipolar,isakov}.  
and the final asymptotic form is analogous to \eqr{eq:Jasymp}
($\Jcal_{\alpha\beta} \propto 1/R^9$ with an oscillating 
angular dependence).
\SAVE{It is the symmetrization of a cubed dipole correlation function.}

%%%%%%%%%%%%%%%%%%%%%%%%%%%%%%%%%%%%
\SAVE{
An equivalent description would have been to call 
$\sigma_i$ a discrete flux along a bond 
(oriented from the even to the odd site) of the honeycomb lattice 
formed by the triangle centers, so \eqr{eq:soft-sum} is a 
zero-divergence constraint.
The coarse-graining maps $\sigma_i$ to a ``polarization''  vector 
field that behaves like a ``magnetic field''~\cite{pseudodipolar,isakov}.  
In $d=2$, the vector potential of that magnetic field
was a potential $\phi_\mu$. In three dimensions,
$\phi_\mu$ cannot be uniquely defined.}
%%%%%%%%%%%%%%%%%%%%%%%%%%%%%%%%%%%%

The crucial difference in $d=3$ is that the discrete (Potts) variables 
also have a ``Coulomb phase'' in place of the ``height representation''
used by Ref.~\onlinecite{huse}.  There exists a coarse-grained 
``flux field'' analogous to $\nabla \hcolor$, but the analog of
$\hcolor$ itself is a vector potential and is not uniquely defined.
The Hamiltonian $\HHPotts$, I conjecture, tends to favor states  
with zero coarse-grained flux, which means it tends to increase
the flux stiffness $K$ of the three-dimensional model.
But, in contrast to two dimensions, in the absence of $\HHPotts$
the system is {\it not} sitting at a critical $K$;
therefore, the tiny increase in $K$ due to the transverse
spin fluctuations cannot drive us into a new phase.
Thus, {\it no long-range order} of the colorings is  expected 
in $d=3$, merely the the pseudodipolar correlations inherent 
to  the Coulomb phase.

\section{Conclusion}
%%% \label{sec:conclusion}
%%% \secc{Conclusion}
%%%%%%%%%%%%%%%%%%%%%%%%%%%%%%%%%%%%%%%%%%%%%
A path has been shown to the elusive long-range order of the
classical kagom\'e antiferromagnet, through a string of mappings
or elimination of degrees of freedom:  ground states to colorings
to chiralities to discrete $\hcolor_\mu$ height representation and 
finally its coarse-grained continuum version. 
Other maps go from all spin deviations, to soft modes $\sigma_i$,
to their height field $\phi_\mu$ or $\phi(\rr)$.
The boldest approximations were 
(i) the perturbation expansion \eqr{eq:Jresult}
of the effective quartic Hamiltonian \eqr{eq:quartic-soft};
this had no controllable small parameter, but it was argued the
terms were numerically small (ii) the variational/decoupling handling of the
quartic ensemble $\{\sigma_i \}$.
In place of the approximations used here, the more elaborate but more 
controlled large-$N$ approach~\cite{canals-garanin,isakov}
(where $N$ is the number of classical spin components) 
looks promising as a formal way to vindicate both approximations.

The philosophy followed here~\cite{henley-HFM00} 
is to obtain an effective Hamiltonian defined for 
{\it arbitrary} spin arrangements, not just
specially symmetric ones (even if that necessitates cruder approximations).
I have previously used the trick of turning the spin configuration 
into a set of coefficients or matrix entries and then expanding in them 
for several systems~\cite{henley87,Chan94,pyrochlore}.
In particular, a related expansion in $\HHthree$ to obtain a 
Hamiltonian of form \eqr{eq:HPotts-pair}
was carried out for the large-$S$ {\it quantum}
Heisenberg antiferromagnet in Ref.~\onlinecite{Chan94}.  

\acknowledgments
This work was supported by NSF grant DMR-0552461.
I thank E. Shender, B. Canals, and M. E. Zhitomirsky for
discussions.


\begin{thebibliography}{99}
\bibitem{FN-not-1to1}
Any 3-coloring can be mapped into chiralities $\{ \eta _i \}$,
not necessarily vice versa; however, if the return map is possible
it is unique [modulo a cyclic permuation of $(ABC)$].
\bibitem{FN-higher-order}
We could also expand \eqr{eq:HHPotts} to the next 
order in $\{ \eta _\alpha \}$ and estimate the coefficients
of four-$\eta$ terms by the same analytic steps used
in this paper to get $\Jcal_{\alpha\beta}$.
\end{thebibliography}
\end{document}